\documentclass[
reprint, 
prd, 
aps, 
nofootinbib, 
showpacs, 
superscriptaddress]{revtex4-2}

\usepackage{graphicx}
\usepackage[usenames,dvipsnames]{xcolor}
\usepackage[utf8]{inputenc}
\usepackage[T1]{fontenc}
\usepackage{anyfontsize}
\usepackage{adjustbox}
\usepackage{placeins}
\usepackage{lipsum}
\usepackage{latexsym}
\usepackage{rotating}
\usepackage{mathtools}
\usepackage{amsmath, amssymb, amsthm, amsfonts}
\usepackage{mathrsfs}
\usepackage{bm}
\usepackage{enumerate}
\usepackage{soul}
\usepackage{natbib}
\usepackage{hyperref}
\usepackage{verbatim}
\usepackage[normalem]{ulem}
\usepackage{cancel}

\hypersetup{colorlinks=true,citecolor=NavyBlue,linkcolor=NavyBlue,urlcolor=NavyBlue}

\newcommand{\2}{\hspace{2mm}}

\newcommand{\h}{\bar{h}}

\numberwithin{equation}{section}
\renewcommand{\theequation}{\arabic{section}.\arabic{equation}}

\begin{document}

\title{Better early than never: \\ A new test for superluminal gravitational wave polarizations}

\author{Kristen Schumacher}
\affiliation{The Grainger College of Engineering, Illinois Center for Advanced Studies of the Universe, Department of Physics, University of Illinois at Urbana-Champaign, Urbana, IL, 61801, USA}

\author{Colm Talbot}
\affiliation{Kavli Institute for Cosmological Physics, The University of Chicago, Chicago, IL 60637, USA}

\author{Daniel E. Holz}
\affiliation{Kavli Institute for Cosmological Physics, The University of Chicago, Chicago, IL 60637, USA}
\affiliation{Department of Physics, University of Chicago, Chicago, IL 60637, USA}
\affiliation{Department of Astronomy \& Astrophysics, The University of Chicago, Chicago, IL 60637, USA}
\affiliation{Enrico Fermi Institute, The University of Chicago, Chicago, IL 60637, USA}

\author{Nicol\'as Yunes}
\affiliation{The Grainger College of Engineering, Illinois Center for Advanced Studies of the Universe, Department of Physics, University of Illinois at Urbana-Champaign, Urbana, IL, 61801, USA}

\date{January 30, 2025}

\begin{abstract}

In some beyond-Einstein theories of gravity, gravitational waves can contain up to six polarizations, which are allowed to propagate at different speeds faster than light.  
These different propagation speeds imply that polarizations generated by the same source will not arrive simultaneously at the detector. 
Current constraints on the speed of propagation of transverse-traceless polarizations, however, indicate that any additional polarizations must arrive with or before the transverse-traceless ones.
We propose a new technique to test for the existence of superluminal, non-transverse-traceless polarizations that arrive in the data before a gravitational-wave observation of transverse-traceless modes. 
We discuss the circumstances in which these non-transverse-traceless polarizations would be detectable and what constraints could be placed if they are not detected. 
To determine whether this new test of general relativity with gravitational wave observations is practical, we outline and address many of the challenges it might face. 
Our arguments lead us to conclude that this new test is not only physically well-motivated but also feasible with current detectors. 
\end{abstract}

\maketitle

\section{Introduction}

The polarizations of gravitational waves (GWs) describe the pattern with which they stretch and squeeze spacetime. 
In general relativity (GR), there are only two transverse and traceless (TT), tensor polarizations, commonly called the plus and cross modes for the pattern they imprint on a circle of test particles when moving perpendicular to it. 
However, an extensive body of literature exists about extensions to GR (see e.g.~\cite{Berti:2015itd, Yunes:2013dva}), some of which predict GWs with more than the two TT tensor polarizations. 
In the most general case, beyond-Einstein gravity predicts GWs with six distinct polarizations: two tensor modes that are TT, and 4 non-tensorial modes (two vector modes, one breathing scalar mode, and one longitudinal scalar mode).
Furthermore, in some of these theories, the different GW polarizations are allowed to propagate at different speeds, all of which can be different from the speed of light. 

Different propagation speeds would result in GWs with different polarizations arriving at different times at the detector, even though they were generated simultaneously by the same source. 
As noted in~\cite{Schumacher:2023jxq}, even a tiny difference in speeds would lead to GW polarizations arriving outside the current standard search window around a trigger event.
Thus, search techniques based on matched-filtering that are used today(e.g.,~\cite{Sathyaprakash:1991mt, Blanchet:2013haa, Usman:2015kfa}), which consider only the simultaneous arrival of different polarizations (if they search for additional polarizations at all), are likely to miss non-tensorial modes entirely.
Similarly, null stream tests~\cite{Guersel:1989th, Chatziioannou:2012rf, Pang:2020pfz, Wong:2021cmp} which implicitly assume that non-tensorial polarizations arrive within their analysis window, are likely to miss non-tensorial modes~\cite{Schumacher:2023jxq}.
In this work, we address this issue by proposing a new method to search for additional polarizations in data, given a detection of TT tensor modes.
Prior observations have already been used to prove that any additional polarizations must arrive earlier than the tensor ones to avoid Cherenkov constraints~\cite{Moore:2001bv, Kiyota:2015dla, LIGOScientific:2017zic, Schumacher:2023jxq}. 
Therefore, we propose starting from already detected events and searching back through earlier data for additional polarizations.

\begin{figure}
    \centering
    \includegraphics[width=\linewidth]{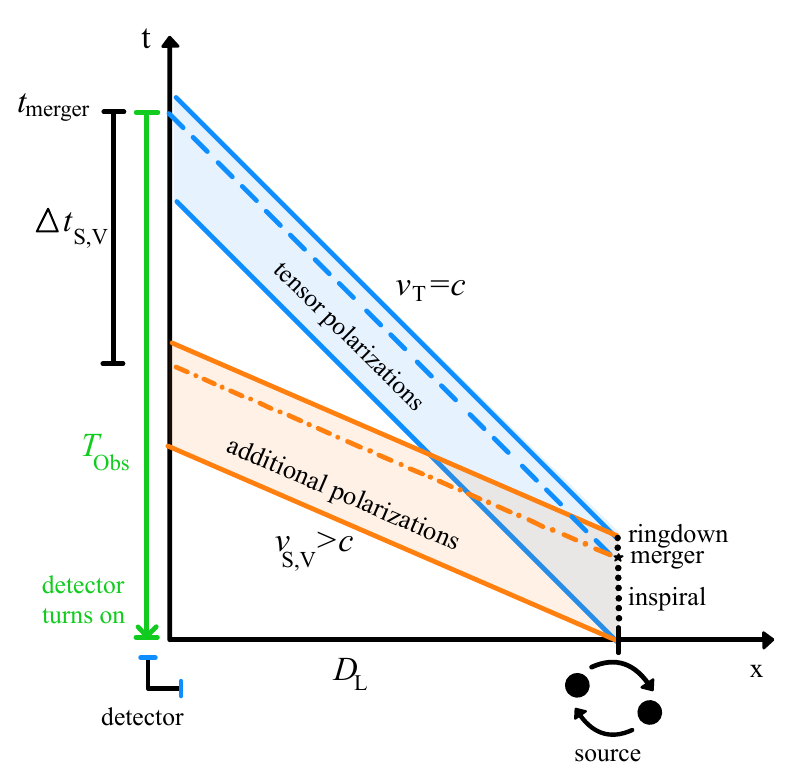}
    \caption{A spacetime cartoon to illustrate our proposed search method. Consider a source of gravitational waves that is a luminosity distance $D_L$ from a detector. The tensor polarizations of these waves (in the blue shaded region) propagate at speed $c$. The portion of the signal containing the merger (between the blue lines) arrives at the detector with peak amplitude at time $t_{\text{merger}} = D_L/c$. Any additional scalar or vector polarizations (between the orange lines) propagate with a speed $v_{S,V} \geq c$, arriving at the detector with peak amplitudes at time $D_L/v_{S,V}$. In this diagram, we assume the non-tensorial polarizations travel significantly faster than the speed of light to exaggerate the effect for visualization purposes. The difference in arrival times between the tensor and the non-tensorial polarizations (see Eq.~\eqref{eqn:time-diff-def}) is $\Delta t_{S,V}$. We propose searching back through the data from $t_{\text{merger}}$ through some $T_{\text{Obs}}$ (in this diagram, $T_{\text{Obs}} = t_{\text{merger}} - t_{\text{observation start}}$) for additional polarizations with an adapted template bank.}
    \label{fig:basic_spacetime}
\end{figure}

Consider for example a GW source at some distance $D_L$ from a detector, as illustrated in the spacetime diagram of Fig.~\ref{fig:basic_spacetime}. 
Suppose we detect the inspiral, merger, and ringdown of the tensor polarizations of this source with our typical matched-filtering techniques using a GR waveform template.
We can determine the distance to the source and the properties of the source (e.g., the mass of the system, the sky location, etc.) from parameter estimation with this data. 
The GW170817 event has already revealed that tensor polarizations propagate effectively at the speed of light $c$ (illustrated in blue in Fig.~\ref{fig:basic_spacetime})~\cite{LIGOScientific:2017zic}. We also know that any additional polarizations must propagate at or faster than the speed of light to avoid Cherenkov constraints~\cite{Moore:2001bv, Kiyota:2015dla} (illustrated in orange in Fig.~\ref{fig:basic_spacetime}). 
Thus, if we start from $t_{\text{merger}}$, the time at which the merger was detected with the tensor polarizations, we can scan back through the data for some observation time, $T_{\text{Obs}}$, with a modified gravity waveform template bank which is sensitive to additional polarizations. 
We can significantly reduce the size of the template bank needed, by performing a targeted search with source parameters informed by the analysis of the tensor polarizations, in much the same way that searches for strongly lensed signals do~\cite{Li:2019osa, McIsaac:2019use, Li:2023zdl}. 
This reduced template bank and the fact that we have limited the amount of data to search over (only before $t_{\text{merger}}$) makes the search dramatically more efficient than a blind search through all data for all additional polarizations.

A direct detection of additional polarizations would of course be clear evidence of new physics beyond GR, but even a non-detection would be informative. 
A non-detection would imply a \textit{lower} bound on the propagation speed of polarizations that have an amplitude measurable by current detectors; the bound is a lower one because the non-detection of additional polarizations by looking back through the data for a time $T_{\rm{Obs}}$ cannot rule out waves that could have arrived even earlier than that, and thus, would have traveled at higher speeds than some lower bound. 
Likewise, a non-detection could also imply an \textit{upper} bound on the amplitude of polarizations with measurable propagation speeds, given the segments of data searched.  
The magnitude of these new constraints, indicating the amount of parameter space that could be ruled out by a non-detection, would depend on the distance to the source and the amount of time searched back through the data. 
Longer observing times and closer\footnote{This may seem counter-intuitive, if one is used to thinking about tests which constrain speeds to be close to $c$. However, in the case of a non-detection lower bound on the speeds, this is not the case, as explained later around Fig.~\ref{fig:spacetime_two_distances}.} sources allow us to place more stringent lower bounds, and thus, rule out a larger portion of propagation speed parameter space.
For instance, in an optimistic case, if no additional polarizations were detected in a year leading up to the detection of tensor polarizations for a source at 50 Mpc, we would be able to place a lower bound on the speed of additional polarizations of $v> (1 + 6 \times 10^{-9}) c$, where $c$ is the speed of light.
Though this lower bound on the speed of propagation may not seem strong, it would be enough to cause year-long gaps between the tensor and any non-tensorial polarizations of far away sources. 
Thus, this provides a clear and concrete motivation for isolated searches for additional polarizations, like the ones carried out (or proposed) in the past
~\cite{LIGOScientific:2020tif, LIGOScientific:2017ycc, LIGOScientific:2018dkp, LIGOScientific:2019fpa, Isi:2017fbj, Isi:2015cva, Isi:2017equ}, comparing data to ``full vector'' or ``full scalar'' cases, which are not as extreme as previously believed. 

The model-agnostic constraints established through a non-detection of additional modes, either on propagation speed or amplitude, could then be mapped into constraints on the coupling constants in specific modified gravity theories. 
Einstein-\ae{}ther gravity~\cite{Jacobson:2004ts} provides one potential candidate for this mapping, as this theory generically predicts additional polarizations that propagate at different speeds.
This theory is part of a class of theories, called Lorentz-violating theories, which are well-motivated by attempts to quantize gravity~\cite{Mattingly:2005re}. 
In fact, Einstein-\ae{}ther gravity is the most general Lorentz-violating theory that can be constructed from a single additional unit vector field and still leads to second-order field equations~\cite{Jacobson:2007veq}.
The current constraints on Einstein-\ae{}ther theory were recently summarized in~\cite{Schumacher:2023cxh}, and in this paper we consider how those constraints might be updated after the test proposed herein. 
Unfortunately, the effect of placing further constraints on propagation speed in this theory is suppressed by coupling constants, which are already tightly constrained by cosmological, solar system, and binary pulsar observations~\cite{Carroll:2004ai, Muller:2005sr, solarSystem2, Gupta:2021vdj}.
Like most model-agnostic constraints, the method proposed in this paper pales in comparison to theory-specific constraints. However, the proposed method allows for the study of a larger set of theories simultaneously and for the possibility of detecting the un-modeled and unexpected.

The model-agnostic test proposed here can be carried out today, with current and near-future data, in spite of minor challenges that can be addressed as more data is collected. 
One such challenge is gaps in the observing time, due to various situations that force the detectors to lose online and coincident duty cycles. 
However, as different tensor polarizations are detected at different times, we show that the loss of duty cycle does not deteriorate the proposed test. 
``Stacking''\footnote{Note that this is slightly different from the typical ``stacking'' of events, which probe the \textit{same} region of parameter space for parameters shared across observations, thus making a constraint tighter by combining multiple measurements.} information from multiple different GW events, which individually allow us to probe different regions of the parameter space, enables the exploration of large portions of the parameter space without any gaps.
We carefully investigate how such data gaps, as well as observing time durations and detector sensitivity, would impact our ability to detect or constrain additional modes. We propose strategies that help mitigate these challenges and show that they can be easily overcome. 

The remainder of this paper is organized as follows. 
Section~\ref{sec:constraining_speeds} discusses constraints on the speed of additional polarizations. 
Section~\ref{sec:current_constraints} reviews the current constraints on the propagation speeds of GW polarizations, while Sec.~\ref{sec:compare_to_tensors} explains how new constraints might be placed by comparing the arrival times (or lack thereof) of additional polarizations to tensor modes from the same event. 
Section~\ref{sec:challenges} considers what challenges we may encounter when carrying out this novel test. 
Section~\ref{sec:observing_time} goes into detail about the observing time and how gaps in the data may affect constraints, an effect that could be mitigated by stacking events. 
Section~\ref{sec:sensitivity} examines how changes in detector sensitivity across runs should be accounted for. 
Section~\ref{sec:magnitude} reviews current constraints on the amplitudes of additional polarizations and discusses how constraints on amplitude and propagation speed may be related (or unrelated) using a specific theory of modified gravity as an example. 
Finally, Sec.~\ref{sec:conclusion} summarizes our results and outlines future work. 
Appendix~\ref{sec:appendix} provides an estimate of the amount of energy carried away by additional polarizations, while Appendix~\ref{sec:independent} considers how new constraints might be placed by comparing arrival times of additional polarizations in different detectors (assuming direct detection). Henceforth, we adopt the following conventions: $G = 1$, but factors of $c$ are explicitly kept for easy comparison between the speeds of different polarizations.

\section{Constraining the speed of additional polarizations}
\label{sec:constraining_speeds}
In this section, we review the current constraints on propagation speed for the different polarizations and conclude that any additional polarizations must have arrived before the correlated tensor polarizations. 
Thus, we propose a new method to search back through the data from the arrival of the tensor modes for any additional polarizations. 
We discuss under what circumstances additional modes might be detectable and what this means for the constraints that can be placed through a non-detection. 

\subsection{Current constraints}
\label{sec:current_constraints}
GWs in modified theories of gravity may contain as many as six polarization modes. 
These include the two tensor polarizations of GR, two additional vector polarizations, and two additional scalar polarizations. 
In general, scalar, vector, and tensor polarizations are allowed to propagate at different constant speeds, $v_S, v_V,$ and $v_T$, respectively. 
This has important implications for detectability since waves generated at the same source that travel at different speeds arrive at the detector at different times. 
As pointed out in~\cite{Schumacher:2023jxq}, even a small difference in speeds can build up over large distances. 
Thus, this effect could be significant for sources that are far away.

Some constraints on the speeds of propagation of GWs already exist. 
If $v_{S, V, T} < c$, then massive particles traveling close to the speed of light, like cosmic rays, would be able to travel faster than GWs, producing gravitational ``Cherenkov''-type radiation, and, thereby, losing energy~\cite{Moore:2001bv, Kiyota:2015dla}. 
However, observations of high-energy cosmic rays rule out such gravitational Cherenkov radiation. 
This implies that $v_{S,V,T} \geq c$~\cite{Moore:2001bv, Kiyota:2015dla}.
Faster-than-light propagation for GW polarizations does not violate causality; it simply introduces a new causal cone~\cite{Geroch:2010da}.
Furthermore, the coincident detection of GWs and a gamma-ray burst from a binary neutron star merger, GW170817 and GRB170817A, established that the tensor polarizations propagate at approximately the speed of light: $v_T \approx c \pm \mathcal{O}(10^{-15})$~\cite{LIGOScientific:2017zic}.
Combining these constraints, one concludes that $v_{S, V} \geq v_{T}$ up to uncertainties of ${\cal{O}}(10^{-15})$, and thus, non-tensorial polarizations must arrive \textit{with} (if $v_{S, V} = v_{T}$) or \textit{before} (if $v_{S, V} > v_{T}$) the tensor polarizations. 

Recent work~\cite{Dong:2023bgt} suggests that, when $v_T = c$, some modified gravity theories have no vector polarization\footnote{For example, this happens to be true in Einstein-\ae{}ther theory, because, if $v_T = c$, one of the coupling constants, $c_\sigma$, must be zero, and if this is so, the amplitude of the vector polarization also vanishes, as one can see from Eq. (3.28) of~\cite{Zhang:2019iim}.}.
The two classes of theories examined in that work, however, were a general metric theory and a scalar-tensor theory, neither of which contained a vector field to source a vector polarization in the first place. 
No proof of a generalization of this claim to vector-tensor or scalar-vector-tensor models (e.g.~the theories discussed in~\cite{Sagi:2010ei, Liang:2022hxd, Rastall:1976uh, Rosen:1971qp, Rosen:1974ua, Lightman:1973kun}) 
exists yet.
Therefore, for the remainder of this paper, we will outline our method for both scalar and vector polarizations. 
If theoretical considerations conclusively rule out vector polarizations in the future, our method would still apply to scalar polarizations.

\subsection{Constraints on the propagation speed of scalar or vector polarizations: Anchoring to a tensor GW observation}
\label{sec:compare_to_tensors}
The LVK collaboration has already detected upwards of 170 events, and this number will only increase as the fourth observing run continues~\cite{KAGRA:2021vkt, LIGO_press_release}.
Analyses of the events in the GWTC-3 catalog suggests that they are consistent with purely tensor polarizations, as predicted by GR~\cite{LIGOScientific:2021sio}. This conclusion was reached through the comparison of the Bayes factors between different models (purely tensor, purely vector, purely scalar, or mixed polarization content)~\cite{LIGOScientific:2021sio}. All models assumed that all polarizations travel at the speed of light, and thus, they could all have caused the data triggers analyzed. This, of course, is not true for a mixed polarization model where the non-tensorial modes are allowed to travel faster than the speed of light, as considered here~\cite{LIGOScientific:2021sio, Schumacher:2023jxq}.
Therefore, this test does not rule out additional polarizations, although it does suggest that the GWs detected so far are likely tensor polarizations.
As an addendum to this analysis, any constraints on the GW dispersion relation~\cite{LIGOScientific:2020tif} apply only to the tensor polarizations.

Since any additional modes must arrive with or before the tensor modes, we here propose using the already detected events to determine which portion of the data to search for additional polarization content.
For example, to probe for scalar or vector polarizations corresponding to the GW170817 event, one would search the data before the trigger time of the detected merger for a signal that has intrinsic 
source parameters consistent with that of the GW170817 event. The source parameters might be slightly biased, because the GR waveform models used to analyze the event do not account for energy being lost through additional polarizations. However, these modifications due to energy loss in other polarizations generally scale with the square of the coupling constant, so they should be suppressed (in the small coupling limit) and can thus be neglected (see Appendix~\ref{sec:appendix} for further details). 

\begin{figure}
    \centering
    \includegraphics[width=\linewidth]{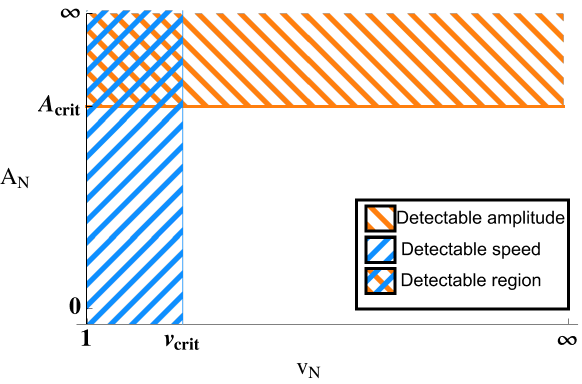}
    \caption{Schematic diagram of the parameter space that could be probed through the non-detection of additional polarizations in a stretch of data prior to a detected tensor signal. Some critical velocity, $v_{\text{crit}}$, exists above which additional polarizations travel so fast that they would have arrived at our detector before it turned on -- we would not be able to measure them. Some critical amplitude, $A_{\text{crit}}$, exists below which current detectors would not be sensitive enough to measure these polarizations. Therefore, the region of detectability is where both speeds and amplitudes are detectable, the overlap region in the upper left-hand corner. Over time, this overlap region would grow larger as $v_{\text{crit}}$ increases with more observing time and $A_{\text{crit}}$ decreases with improved detector sensitivity.}
    \label{fig:excluded_param}
\end{figure}

One can imagine two different reasons to explain why additional modes would not have been observed before the detected tensor polarizations, beyond the null-hypothesis that they do not exist.
First, the additional modes may not have had a large enough amplitude for current instruments to detect them.
Second, they may have arrived before the stretch of data searched, because of their propagation speed (recall $v_{S, V} \geq c$). 
Thus, a non-detection of additional modes in earlier data would allow us to place an upper bound on the amplitude of additional polarizations that propagate below a particular speed, and a new lower bound on the speed of propagation for polarizations with a larger amplitude. 
The different regions of parameter space that could be excluded are shown schematically in Fig~\ref{fig:excluded_param}.

\begin{figure}
    \centering
    \includegraphics[width=\linewidth]{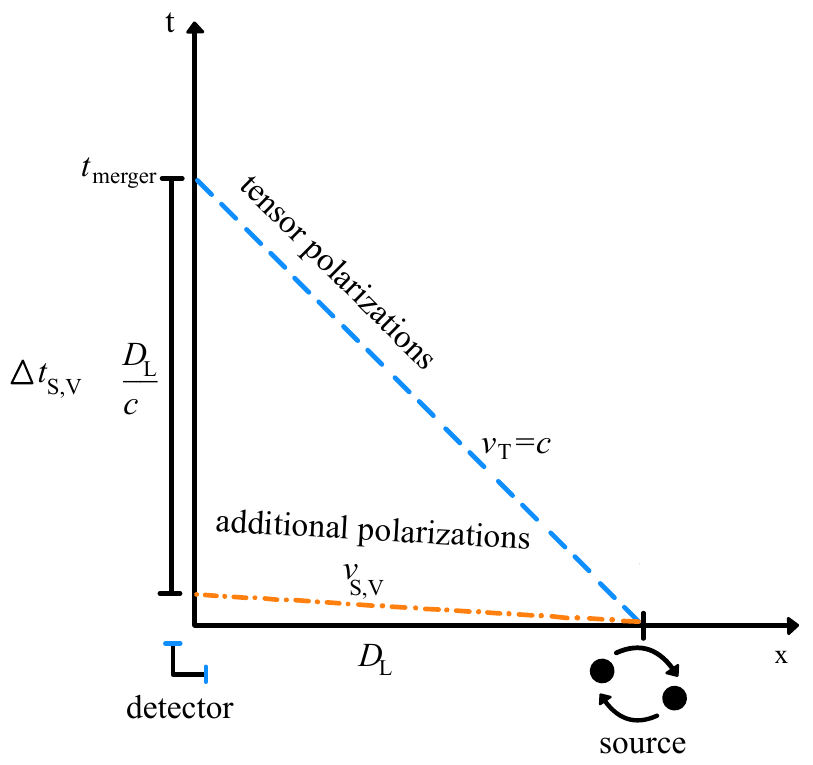}
    \caption{In the limit that the speed of  additional polarizations goes to infinity, their worldline (dot-dashed orange line) approaches a  horizontal line, meaning they arrive almost instantaneously with the merger. 
    Meanwhile, it will still take the tensor polarizations (dashed blue line) a time $D_L/c$ to reach the detector. Therefore, $\Delta t_{S,V} \rightarrow D_L/c$. The alternative polarizations cannot arrive \textit{before} the GW event. Thus, the difference in arrival times cannot be greater than the time it took for the tensor polarizations to reach the detector. This is why $\Delta t_N$ remains finite in this limit. }
    \label{fig:limit_diagram}
\end{figure}

Let us consider non-tensorial polarizations with amplitudes large enough to be detectable with current instruments. 
If these additional polarizations are not found, the length of the data searched sets the minimum possible difference in arrival times (between the non-tensorial and the tensorial polarizations). 
It is a matter of simple algebra to show that~\cite{Schumacher:2023jxq} 
\begin{align}
    \Delta t_{N} &= t_T - t_N = \frac{D_L}{c} - \frac{D_L}{v_N},
    \label{eqn:time-diff-def}
\end{align}
where $t_{T,N}$ are the travel times of the tensor and the non-tensorial polarizations ($N \in \{S, V\}$), $v_N$ is the propagation speed of the non-tensorial polarization, $\Delta t_N$ is the difference in arrival times (between the non-tensorial and tensorial polarizations), and $D_L$ is the luminosity distance between the source and the detector, which is the same for all polarizations from the same event. 
In the limit that $v_N \rightarrow c$, the tensor and non-tensorial polarizations propagate at the same speed, and so they arrive at the same time: $\Delta t_N \rightarrow 0$, as expected. 
As $v_N \rightarrow \infty$, the time that it takes for additional polarizations to arrive in the detector after they are emitted by the source goes to zero, $t_N \to 0$, while the amount of time for the tensor polarizations to arrive is still $t_T = D_L/c$ (Fig.~\ref{fig:limit_diagram} illustrates this with a spacetime diagram). 
Note that alternative polarizations cannot arrive before they are generated at the source. 
Thus, it makes sense that the difference in arrival times is finite even as $v_N \rightarrow \infty$, and that the maximum possible difference in arrival times is the time it takes for the tensor polarizations to propagate between the source and the detector.

We can rearrange the expression of Eq.~\eqref{eqn:time-diff-def} to more conveniently illustrate the differences in speeds as
\begin{equation}
    \frac{v_N}{c} = \frac{1}{1 - \frac{c \Delta t_{N}}{D_L}} = 1 + \delta v_N,
    \label{eqn:speed_diff}
\end{equation}
if we define the percent difference, $\delta v_N$, as
\begin{equation}
    \delta v_N \equiv \frac{v_N}{c}-1 = \frac{c\Delta t_{N}}{D_L - c\Delta t_{N}}.
    \label{eqn:delta_N}
\end{equation}
With this definition, the condition $v_{S, V} \geq c$ implies that $\delta v_N$ is positive\footnote{ $\delta v_N \geq 0$ again implies that $0 \leq \Delta t_N \leq D_L/c$, which means that the difference in arrival times cannot be greater than the time it took for the tensor polarizations to reach the detector. In other words, the alternative polarizations cannot arrive \textit{before} the GW event. The case $\Delta t_N = D_L/c$ would require $v_N \rightarrow \infty$.}.
These expressions show that as $\Delta t_{N}$ increases from zero and approaches $D_L/c$, $\delta v_N$ increases (recall that $\Delta t_{N}$ cannot be larger than $t_T = D_L/c$). 
Conversely, as $D_L$ decreases, forcing $\Delta t_N$ to also decrease, $\delta v_N$ increases. 
Therefore, \textit{looking at longer stretches of data or closer events would allow us to place more stringent constraints.} 

\begin{figure}
    \centering
    \includegraphics[width=\linewidth]{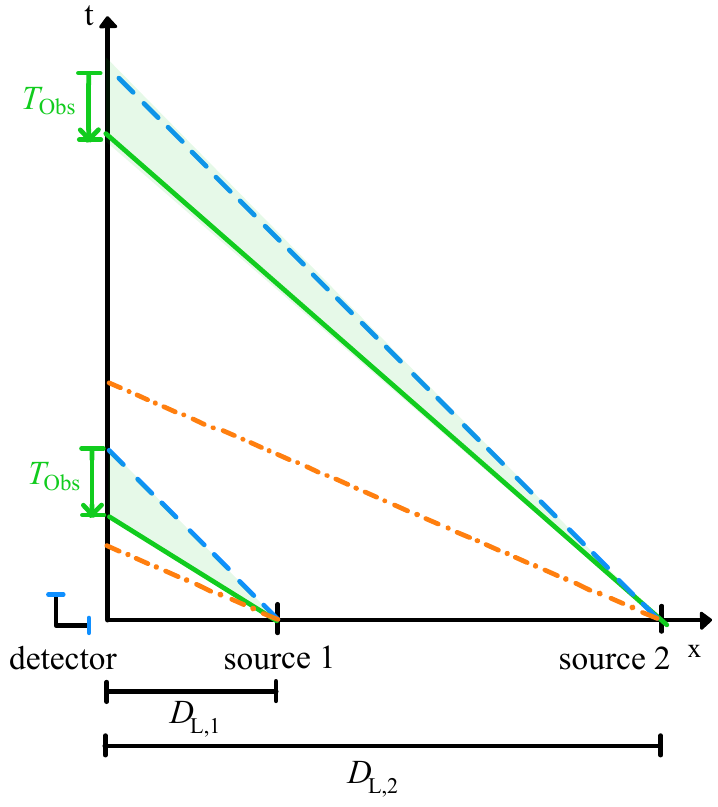}
    \caption{Consider two sources at different distances from the detector. Both emit tensor polarizations propagating at speed $c$ (dashed blue lines) and some additional polarizations propagating at the same speed $v_N > c$ (dot-dashed orange lines). From Eq.~\eqref{eqn:time-diff-def} and from this diagram, we can see that as $D_L$ increases, so does the difference in arrival times between the tensor polarizations and any additional polarizations. Thus, the same amount of observing time searched back from the merger of both events will rule out different regions of the parameter space. The same $T_{\text{Obs}}$ for the closer source 1 will rule out a larger swath of angles (shaded in green) than $T_{\text{Obs}}$ for the more distant source 2. This provides an intuitive understanding of why closer sources are better for ruling out larger regions of parameter space, and hence place better constraints in the event of a non-detection.}
    \label{fig:spacetime_two_distances}
\end{figure}

That searching over longer stretches of data leads to a more stringent constraint on the propagation speed of non-tensorial modes is indeed intuitive, but the same is not necessarily true for sources that are nearby. 
Why would it be that closer signals allow for more stringent constraints on the propagation speed? 
After all, the closer the sources, the less time the signal has to propagate through spacetime, and the smaller the time difference that builds up between the tensorial and the non-tensorial polarizations, as one can see from Eq.~\eqref{eqn:time-diff-def}. 
The reason the constraint improves is that this smaller time difference build up is an advantage when ruling out portions of the parameter space. 
In the case of a non-detection, additional polarizations coming from a closer source had to be traveling much faster to arrive outside the window of time searched. 
This concept is illustrated in Fig.~\ref{fig:spacetime_two_distances}, for GW sources at two different distances. 
In this diagram, both sources emit tensor polarizations propagating at one speed and additional polarizations propagating at another.
Suppose that for both events we  search back through the data for the \textit{same} amount of time, $T_{\text{Obs}}$. 
If we trace a line from the GW event to the earliest time searched in $T_{\text{Obs}}$, we can see the range of angles (and hence speeds on this spacetime diagram) that would be ruled out by a non-detection with this observation. 
We can see that for a closer event, the same observing time allows us to rule out a much larger range of propagation speeds. 

Now, let us instead consider non-tensorial polarizations with propagation speeds detectable by current observations, and examine what a non-detection would tell us about their amplitudes. 
If these additional polarizations are not found, the minimum amplitude that can be detected becomes a new upper bound on their amplitude at the detector.
However, the amplitude of GWs decreases with the distance between the source and the detector, like $1/D_L$. 
Thus, \textit{closer events again allow for more stringent constraints}.
If we want to stack constraints from multiple events, we should consider what this constraint means for their amplitude at the source, or we should consider the ratio of amplitudes between non-tensorial polarizations and the tensor ones (as both should fall off with distance in the same way). 

In the sections that follow, we will describe the possibility of placing constraints on propagation speed for polarizations with detectable amplitudes (unless otherwise specified). 

\section{Addressing challenges to constraints}
\label{sec:challenges}
In this section, we explore and attempt to address some of the limitations of the model-agnostic test  proposed in the previous section. 
We consider how the observing time and the sensitivity of the detectors impact our ability to detect additional modes. 
From there, we can determine how they would impact the possible constraints. 
We can mitigate some of these challenges by continuing to observe, improving detector sensitivity, and stacking events. 
Finally, we consider the current constraints on the magnitude of additional polarizations to ensure that detectable polarizations are not yet ruled out. 
In light of these discussions, we conclude that such a search may yield constraints on propagation speed.

\subsection{Observing time}
\label{sec:observing_time}
Our method's ability to constrain the propagation speed of GWs is limited by the total amount of time that detectors have been observing GWs. 
Recall from Sec.~\ref{sec:compare_to_tensors}, as the difference in arrival times, $\Delta t_N$, increases, so does the difference in speeds, $\delta v_N$. 
Thus, the more data available before a particular set of tensor polarizations arrives at the detectors, the greater the region of parameter space that can be ruled out from a non-detection, and the greater the constraint that can be placed on the propagation speeds of additional polarizations. 
However, GW observations are a relatively young field. 
The first detection was in 2015~\cite{LIGOScientific:2016dsl}. 
The youth of the data limits the constraint on the propagation speeds of non-tensorial modes that we can currently establish with this method. 

\begin{figure*}
    \centering
    \includegraphics[width=.49\linewidth]{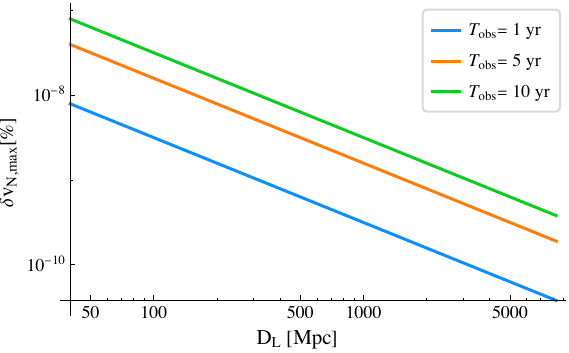}
    \includegraphics[width=.49\linewidth]{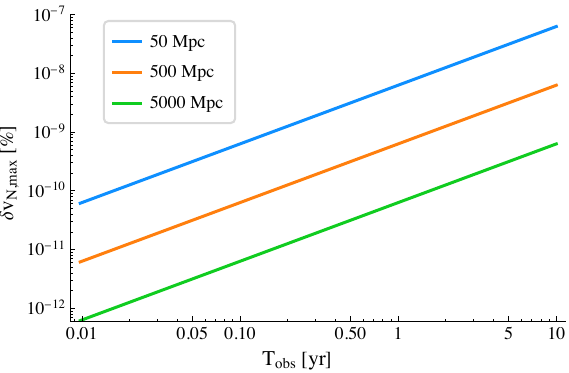}
    \caption{Left: A plot of the maximum observable percent difference in propagation speed, $\delta v_{N, \text{max}} = {v_N}/{c} -1$, as a function of the distance between the source and the detector, for three different observing times. This represents the best-case constraint; the difference in arrival times is equal to the observing time ($\Delta t_{N,\text{max}} = T_{\text{Obs}}$). Therefore, the regions below these lines would be ruled out by a non-detection of additional modes at smaller intervals ($\Delta t_{N} < T_{\text{Obs}}$). Right: A plot of the maximum possible percent difference in speed that could be observed, $\delta_{N,\text{max}}$, as a function of observing time for sources at three different distances. Again, this is the maximum possible difference, so a continuous search would rule out smaller differences.}
    \label{fig:maxDelta_distance}
\end{figure*}

To find the best-case constraint possible at any given time, we need to compute the maximum possible difference in speeds that \textit{could} be detected to date, $\delta v_{N,\text{max}}$.
To do so, we set the maximum observable difference in arrival times to the amount of time observed for, 
\begin{equation}
    \Delta t_{N, \text{max}} = T_{\text{Obs}} = t_{\text{current}} - t_{\text{observation start}}
\end{equation} 
and consider a range of possible distances to the source in Eq.~\eqref{eqn:delta_N}. 
We can project how constraints will improve with increased observing time by changing $t_{\text{current}}$.
The left panel of Fig.~\ref{fig:maxDelta_distance} shows what $\delta v_{N, \text{max}}$ would be after 1, 5, and 10 years of continuous observation, as a function of distance\footnote{The range of $D_L$ was chosen to be $[40, 8280]$ Mpc because these are the closest and farthest events detected to date at the writing of this paper~\cite{LIGOScientific:2017vwq, LIGOScientific:2021usb}.}.
Likewise, we can plot $\delta v_{N, \text{max}}$ as a function of observing time for different distances, as shown in the right panel of Fig.~\ref{fig:maxDelta_distance}.
Note that because $\delta v_{N,\text{max}}$ is the maximum observable difference in speeds, all smaller differences should also be observable with continuous observation and hence could be ruled out by a non-detection of additional modes.
From both plots, it is apparent that the best-case constraint improves with closer events and longer observing times. 
Therefore, though the total amount of observing time limits the maximum constraint, this improves the longer the detectors are on.

The best-case constraints studied above, however, assume an ideal situation in which the detectors consistently collect data from the moment they were first turned on. 
The reality is much more complicated. 
Ground-based GW detectors have been collecting data during a number of planned observing runs, commonly denoted O\# (e.g., O1, O3a), and updated in between (calendar of observing runs available at~\cite{ligo_ORuns_forecast}). 
Therefore, there are several segments of time during which GWs could not be measured. 
If additional polarizations arrived in these gaps between observing runs, we would not have detected them. 
Thus, it seems like instead of being able to place an absolute lower bound on propagation speeds, the best we can hope to do is to rule out a range of propagation speeds.
We could avoid this issue by considering the data from only one observing run at a time. However, that severely limits the magnitude of the constraint that can be placed. 
A better way to address this issue is to consider many different events across different observing runs so that, eventually, all of the regions in parameter space that one particular event cannot rule out are excluded by another. 

\begin{figure*}
    \centering
    \includegraphics[width=.7\linewidth]{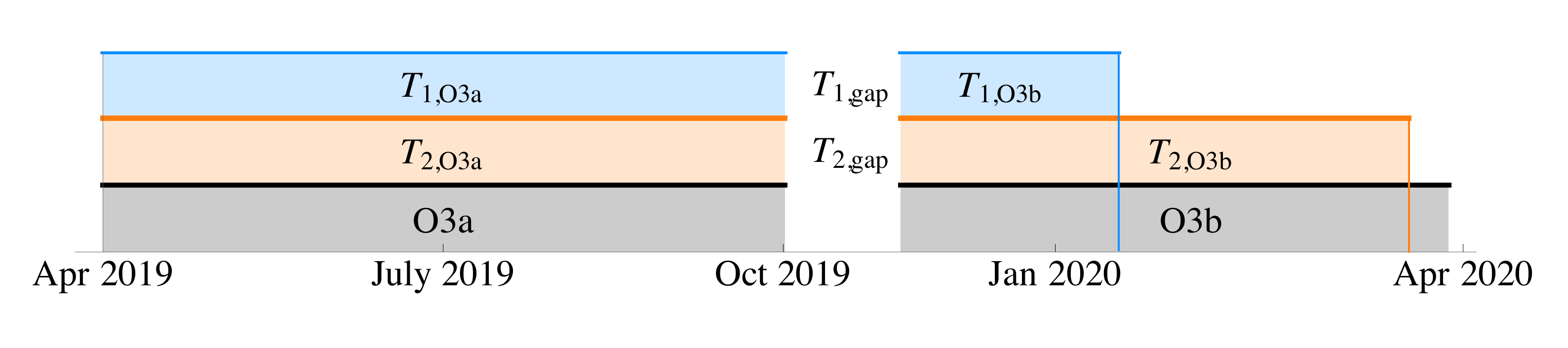}
    \includegraphics[width=.7\linewidth]{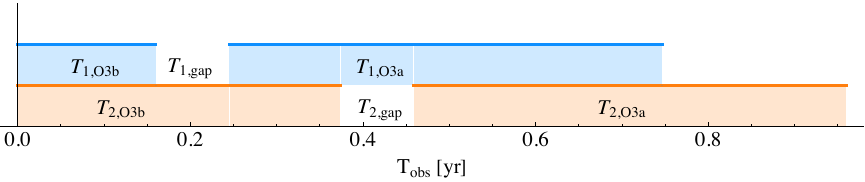}
    \caption{Top: A calendar of searchable data for two hypothetical events with tensor polarizations arriving at different times in O3b. Event 1 arrives earlier (marked by a blue vertical line), and Event 2 arrives later (marked by an orange vertical line). From this calendar, we can see that the length of observing time for both events is the same in O3a and across the gap, but that $T_{1,\text{O3b}} < T_{2,\text{O3b}}$ because event 1 arrived earlier. Bottom: A timeline of the total observing time when looking back from the arrival of the tensor polarizations. Because $T_{1,\text{O3b}} < T_{2,\text{O3b}}$, the gaps in the data do not line up for both events. This has important implications for the regions of $\delta v_{N}$ that can be ruled out by stacking the events (as shown in Fig.~\ref{fig:Delta_gapped_diff_time}).}
    \label{fig:gapped_calendar}
\end{figure*}

To demonstrate how this would work schematically, let us consider a hypothetical scenario in which we change only one variable at a time. 
Suppose that tensor polarizations from two events at the same distance (50 Mpcs) from the detector were observed at different times during the second half of the third observing run, O3b. 
Starting from the arrival of these tensor polarizations, we can search back through earlier data for additional modes. 
The top panel of Fig.~\ref{fig:gapped_calendar} presents a calendar of the different intervals of data that can be searched for each event. $T_{i, \text{O3b}}$ is the time interval in O3b, $T_{i, \text{gap}}$ represents the gap between observing runs when the detector was turned off, and $T_{i, \text{O3a}}$ is the time interval in O3a for $i=1,2$. 
Note that $T_{1,\text{O3b}} < T_{2,\text{O3b}}$ because event 1 arrived earlier in O3b than event 2. 
Meanwhile, the duration of the gap and O3a was the same for both events, so $T_{1,\text{gap}} = T_{2,\text{gap}}$ and $T_{1,\text{O3a}} = T_{2,\text{O3a}}$.
However, if we look at the total observing time we can search for each event, $T_{1, \text{total}} < T_{2, \text{total}}$, and we can see from the bottom panel of Fig.~\ref{fig:gapped_calendar} that because $T_{1,\text{O3b}}<T_{2,\text{O3b}}$, searching backward from the events the gaps do not line up. 
Thus, the two different events can probe different regions of $\delta v_N$.

\begin{figure}
    \centering
    \includegraphics[width=\linewidth]{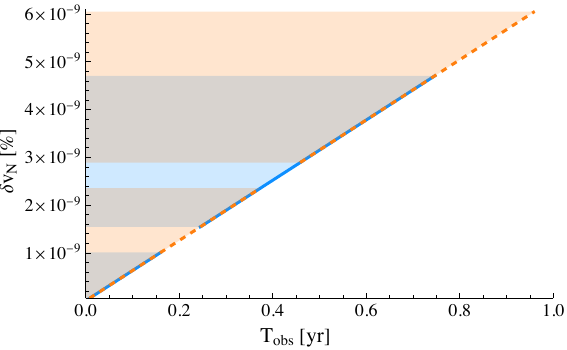}
    \includegraphics[width=\linewidth]{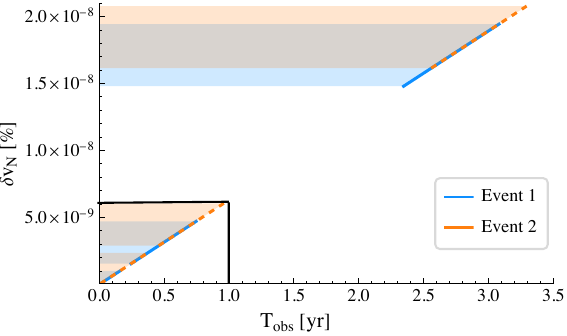}
    \caption{Both panels show the regions of $\delta v_N$ that could be excluded by two events at the same distance of 50 Mpc arriving at different times in O3b (blue and orange shadings represent regions excluded by Events 1 and 2, respectively), and the top panel shows a smaller portion of the bottom plot in greater detail. The top panel shows the regions that a non-detection of additional polarizations in O3a and O3b could exclude. Observe that, although there is a gap in the data for each individual event, these gaps overlap in such a way as to give a continuous region of possible detection/exclusion (see Fig.~\ref{fig:gapped_calendar} for a timeline). The bottom panel shows the regions that a non-detection of additional polarizations in O2, O3a, and O3b could exclude. Here we can see that the much larger gap between O2 and O3a data is made narrower by overlapping data from the two events, but not eliminated completely. The inclusion of more than two events would shrink the gaps further, and, with enough events, they would be eliminated. }
    \label{fig:Delta_gapped_diff_time}
\end{figure}

The shading in the top panel of Fig.~\ref{fig:Delta_gapped_diff_time} shows the different regions of $\delta v_N$ that each event could exclude if additional modes were not detected. 
From this plot, we can see that the gap in observing time between O3a and O3b would not lead to a gap in the measured region of $\delta v_N$ as long as we were able to use events that occurred at different times. 
However, the gap between O3a and O3b was a short time interval, only a month. So, how does this strategy work for a larger gap? 
If we were to include O2 in this search, then we have a much larger gap in observing time between O2 and O3a, about 1.5 years. Making the same plot, we can see from the bottom panel of Fig.~\ref{fig:Delta_gapped_diff_time} that the gap between O3a and O2 data is made narrower by overlapping data from the two events but not eliminated. 
Hence, we should consider another way to fill in such observing gaps.

\begin{figure}
    \centering
    \includegraphics[width=\linewidth]{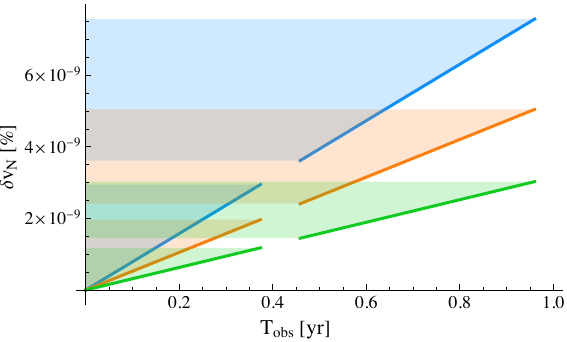}
    \includegraphics[width=\linewidth]{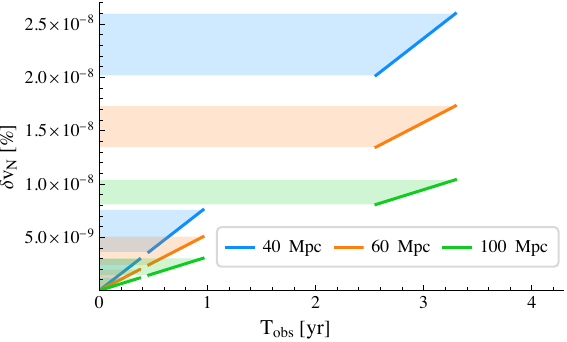}
    \caption{Both panels show the regions (shaded) of $\delta v_N$ that could be excluded by three events at different distances (40, 60, and 100 Mpcs respectively) arriving at the same time in O3b, and the top panel shows a smaller portion of the bottom plot in greater detail. The top panel shows the regions that a non-detection of additional polarizations in O3b and O3a could exclude. Observe that, although there is a gap in the data for each individual event (and the gaps line up on this plot), the different slope for each event leads to an overlap in $\delta v_N$, leading to a continuous region of possible detection/exclusion. The bottom panel shows the regions that a non-detection of additional polarizations in O2, O3a, and O3b could exclude. Observe that the much larger gap between O3a and O2 data is filled in by data from the three events but not eliminated completely. The inclusion of more events would indeed eliminate the gap altogether.}
    \label{fig:gapped_time_diff_distances}
\end{figure}

Consider then that tensor polarizations from three events at different distances from the detector ($40, 60$, and $100$ Mpc respectively) are observed simultaneously. 
In this case, if we search the data back from the arrival of the tensor polarizations for additional modes, the gaps in observing time would line up on the $T_{\text{Obs}}$ axis. 
However, because $\delta v_N$ also depends on $D_L$, the slope of the line on a plot of $\delta v_N(T_{\text{Obs}})$ is different. 
The top panel of Fig.~\ref{fig:gapped_time_diff_distances} shows how this affects the regions of $\delta v_N$ that can be excluded.
Once again, the gap in observing between O3a and O3b does not lead to any gap in the region of $\delta v_N$ that could potentially be excluded. 
However, the larger gap between O2 and O3a data, visible in the bottom panel of Fig.~\ref{fig:gapped_time_diff_distances}, leads to a gap in the observable $\delta v_N$. 
The region of $\delta v_N$ that could not be measured is made much smaller but not completely eliminated by combining these three events. 

\begin{sidewaysfigure}
    \centering
    \includegraphics[width=\linewidth]{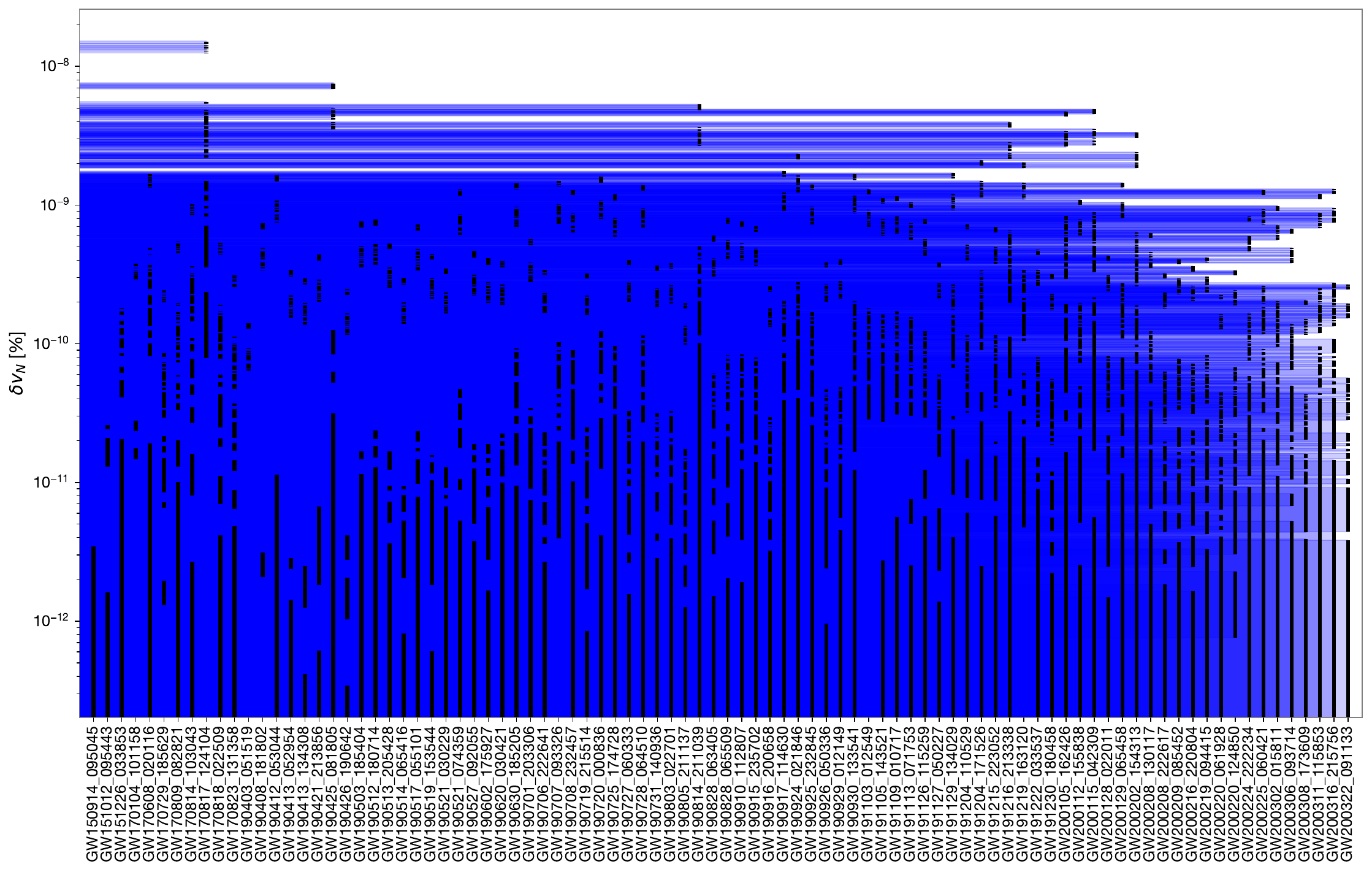}
    \caption{The regions (shaded) of $\delta v_N$ that could be excluded by a non-detection of additional polarizations for real GW events detected in observing runs O1 - O3. The black lines correspond to the $\delta v_N$'s that can be ruled out for each individual event. Observe that there are a few events that lead to the best constraints, namely GW170817\_124104, GW190425\_081805, GW190814\_211039, GW191216\_213338, and GW200115\_042309. Therefore, searches for non-tensorial polarizations should start with these events. }
    \label{fig:real_data_possible_constraints}
\end{sidewaysfigure}

From these two examples, we can see how combining data from many different events that arrive at different times or from different distances can solve the problem of gaps in the observing time\footnote{Note that this ``stacking'' of events would not work for a theory where the propagation speed was expected to be different for different events (e.g. if the speed of propagation depended on the distance at which waves were generated).}.
This technique of combining data from different events can also mitigate gaps in observation within an observing run. 
Even within an observing run, there are periods when the detector is off-line for one reason or another, or when noise or glitches obscure the data.
Any less-than-optimal detection capabilities are quantified by the duty cycle, the fraction of the total run duration during which the instruments were observing~\cite{KAGRA:2021vkt}. 
For example, \cite{KAGRA:2021vkt} reports that during O3b, the duty cycle was 79\% for advanced LIGO Hanford, 79\% for advanced LIGO Livingston, and 76\% for Virgo. All three detectors were running simultaneously only 51\% of the time, but at least one detector was observing 96.6\% of the time. 
This creates further challenges for our proposed technique because, to properly exclude regions of the parameter space, one would have to keep careful track of when each detector was offline and determine how that impacts observable speeds. 
We can see how this would work for real events detected across O1 -- O3 in Fig.~\ref{fig:real_data_possible_constraints}.
Here, we have collected all segments of data in which at least one detector was turned on and calculated what a non-detection of additional polarizations in that segment would imply for constraints on $\delta v_N$ for each GW event. 
We can use this plot to determine which GW events would allow us to rule out the largest regions of $\delta v_N$, if one were to actually perform this search (e.g., GW170817\_124104, GW190425\_081805, GW190814\_211039, GW191216\_213338, and GW200115\_042309 would allow us to rule out larger values of $\delta v_N$).
Thus, we can again see that if one were to repeat this test multiple times with different events over many different observing runs, many gaps in observation would become filled in. 

\subsection{Detector sensitivity}
\label{sec:sensitivity}
A matter to be cautious of when searching for additional modes across multiple different observing runs is the change in the detectors' sensitivity. 
With each new observing run, the sensitivity of the detectors increases, allowing us to detect more distant sources. 
This also means that the detectors become less sensitive as we look back through the data from the arrival of a particular tensor polarization to earlier observing runs. 
Thus, it is possible that additional modes that are detectable in a later run may not be observable in an earlier run. 

We can counter this problem by restricting our search for additional modes to a particular observing run. 
However, as previously discussed, that can limit the regime of $\delta v_N$ that is measurable/constrainable. 
Another solution is to search for additional modes with a lower signal-to-noise ratio (SNR) threshold. 
For example, targeted sub-threshold searches, such as those that look for demagnified, lensed GW signals~\cite{Li:2023zdl}, require a detector SNR $\rho_{\text{det}} > 4$ in at least one detector, as compared to the traditional SNR threshold\footnote{This SNR threshold of 8 is often used as a heuristic, so we quote it here to give a point of comparison. In reality,  SNR alone is not a sufficient detection statistic~\cite{Usman:2015kfa}, and the optimal SNR threshold can vary for different source parameters~\cite{Essick:2023toz}.} of 8~\cite{KAGRA:2013rdx}. 
Thus, there is precedence in the literature for using sub-threshold searches when some of the source parameters have already been determined from a super-threshold event~\cite{Li:2019osa, McIsaac:2019use, Li:2023zdl}.

A third way to counter this problem might be to consider only events close enough to Earth to have been detected in any of the observing runs. 
Yet, as pointed out in the previous section, we might need events at multiple distances to examine a larger (or all of the) $\delta v_N$ parameter space. 
Fortunately, because of how quickly the slope changes and the size of the gaps between observing runs, we may not need a very large range of distances (e.g., between 40 and 100 Mpc is almost enough to remove the gap between O2 and O3a if enough events were detected in this range, Fig.~\ref{fig:gapped_time_diff_distances}).
Furthermore, as demonstrated in Secs.~\ref{sec:compare_to_tensors} and~\ref{sec:observing_time}, closer events give better constraints on $\delta v_N$ (and on amplitude when $\delta v_N$ is measurable). 
Therefore, the issue of being unable to detect additional polarizations for events at very large distances is not necessarily detrimental to constraining propagation speeds (or the amplitude) of non-tensorial modes. 

\subsection{Magnitude of additional polarizations}
\label{sec:magnitude}
Our method to constrain the propagation speed of additional polarizations through a non-detection applies only to polarizations with a large enough non-tensorial amplitude to be detectable (recall Fig.~\ref{fig:excluded_param}). 
Our proposed technique would be irrelevant if current constraints already ruled out detectable amplitudes of additional modes. 
We must therefore consider what constraints exist on the amplitude of additional polarizations. 

Several studies have searched for a GW background (GWB) arising from non-tensorial modes and have placed upper limits on how large that background can be~\cite{Romano:2016dpx, Callister:2017ocg, LIGOScientific:2018czr, LIGOScientific:2019gaw, KAGRA:2021kbb, NANOGrav:2021ini, NANOGrav:2023ygs},
which puts an upper limit on how large the amplitude of alternative polarizations can be.
According to ~\cite{KAGRA:2021kbb}, the upper limits are 
\begin{subequations}
\begin{align}
    \Omega^{(S)}_{GW} &\leq 2.1 \times 10^{-8}, \\
    \Omega^{(V)}_{GW} &\leq 7.9 \times 10^{-9}, \\
    \Omega^{(T)}_{GW} &\leq 6.4 \times 10^{-9},
\end{align}
\end{subequations}
for scalar, vector, and tensor polarizations, respectively. 
\onecolumngrid

\twocolumngrid
\noindent Given that the tensor polarization has the most stringent constraint and tensor polarizations are detectable, these constraints on the GWB do not imply that alternative polarizations would be any more difficult to detect than tensor polarizations. 

A different study~\cite{Takeda:2021hgo} placed constraints on the amplitude of scalar polarizations in a scalar-tensor theory with a direct search for additional polarizations around the merger events GW170814 and GW170817.
However, that analysis assumed that the speed of propagation was the same for the tensor and scalar polarizations. 
Thus, the constraints of that paper do not apply to the theories considered here, as their method would have missed any polarizations that arrived outside the window of data they considered (4 seconds of data for GW170814 and 128 seconds of data for GW170817). 
Similarly, another study~\cite{Chatziioannou:2021mij}, carried out a theory-independent search for additional polarizations, but they considered only the 4 seconds of data around the merger event GW190521, so again their upper limits do not apply to polarizations that propagate with different speeds. 

Another method to place a constraint on the amplitude of additional polarizations is to examine the maximum possible energy loss due to the emission of additional polarization modes, and then to relate this to a constraint on the amplitudes, as was recently done in~\cite{Takeda:2023mhl}.
Their analysis was limited to the case where only one additional scalar mode is present, so they focused on scalar polarizations. 
Thus, any constraints placed by this study do not apply to vector polarizations. 
Furthermore, the authors assumed that the propagation speed of scalar polarizations is very close to the speed of light (which we did not assume here). 
Finally, their main result, that scalar modes should be suppressed by a factor of $\mathcal{O}(10^{-2.5})$ relative to tensor modes, can be avoided if a certain relation between the breathing and longitudinal polarizations holds. 
Given all of these reasons, we do not think the results of~\cite{Takeda:2023mhl} should deter us from the search proposed here. 

Another potential hurdle to detectability is that the relative difference in propagation speed with respect to the speed of light may not be independent of the amplitude of non-tensorial polarizations, since, in a given particular theory, both should be related to the magnitude of the coupling constants of that theory.
If true, this would reduce detectability prospects, as it seems to imply that polarizations that arrive close to the tensor polarizations must have had smaller coupling constants, leading to non-detectable amplitudes. 
However, constraining the speed of non-tensorial polarizations \textit{does not} necessarily constrain their amplitude if there are multiple coupling constants in the theory, as is the case with Einstein-\ae{}ther theory.
Similarly, if there are multiple coupling constants, constraints on the non-tensorial speeds may have significant or little impact on coupling constant constraints, depending on how they enter these speeds. This relation must be determined on a case-by-case basis, in specific theories.

We will examine Einstein-\ae{}ther theory as a specific example. 
In this theory, there are four dimensionless coupling constants, $\{c_a, c_\theta, c_\omega, c_\sigma\}$\footnote{Sometimes, the theory is parameterized in terms of alternative coupling constants, $\{ c_1, c_2, c_3, c_4\}$. See~\cite{Jacobson:2013xta} for the conversion between these two sets of couplings.}~\cite{Jacobson:2013xta}. 
The speeds of the different polarizations are related to these coupling constants through~\cite{Jacobson:2004ts}
\begin{subequations}
\begin{align}
\frac{v_T^2}{c^2} &= \frac{1}{1 - c_\sigma} \label{eqn:tensorSpeed}, \\
\frac{v_V^2}{c^2} &= \frac{c_\sigma + c_\omega - c_\sigma c_\omega}{2c_a (1 - c_\sigma)}, \label{eqn:vectorSpeed} \\
\frac{v_S^2}{c^2} &= \frac{(c_\theta + 2c_\sigma)(1 - c_a/2)}{3c_a(1 - c_\sigma)(1 + c_\theta/2)}. \label{eqn:scalarSpeed}
\end{align}%
\label{eqn:polarization_speeds}%
\end{subequations}%
Let us first consider the tensor polarizations in this theory. Because the speed of the tensor polarizations only depends on $c_\sigma$, any constraints on $v_T/c$ can be easily translated into constraints on $c_\sigma$. This is why the constraints from the GW170817 and GRB170817A event (as described in Sec.~\ref{sec:current_constraints}), which require $v_T \approx c \pm \mathcal{O}(10^{-15})$\cite{LIGOScientific:2017zic}, imply the stringent constraint $c_\sigma \approx \mathcal{O}(10^{-15})$.
This is a clear example in which a constraint on the speed of propagation of a mode (in this case, the tensor modes) places a stringent constraint on the coupling constants of the theory. 

Now, let us consider the hypothetical case where we place a constraint on the speed of the vector polarization, $v_V/c \geq 1 + \delta v_V$, where $\delta v_V$ is whatever difference in speeds between the tensor and vector modes can be ruled out by this constraint. Given the tight constraint on $c_\sigma$ due to the coincident GW and GRB observation, the speeds of different polarizations in Einstein-\ae{}ther theory are effectively
\begin{subequations}
\begin{align}
\frac{v_T^2}{c^2} &= 1 \label{eqn:tensorSpeedUpdated}, \\
\frac{v_V^2}{c^2} &= \frac{ c_\omega }{2c_a }, \label{eqn:vectorSpeedUpdated} \\
\frac{v_S^2}{c^2} &= \frac{c_\theta (1 - c_a/2)}{3c_a(1 + c_\theta/2)}\,,
\label{eqn:scalarSpeedUpdated}
\end{align}%
\label{eqn:polarization_speeds_updated}%
\end{subequations}
setting $c_\sigma = 0$ for simplicity. Then,
\begin{align}
    \frac{v_V^2}{c^2} &\geq 1 + (2\delta v_V + \delta v_V^2)\,,
\end{align}
to leading order in the couplings, and thus
\begin{align}
    c_\omega &\geq 2c_a \left[1 + (2\delta v_V + \delta v_V^2)\right]\,.
\end{align}
The constraint on $c_\omega$ is suppressed by $c_a$, which is already small due to other constraints (see~\cite{Schumacher:2023cxh} for a summary of current constraints on this theory). Therefore, we now have an example where constraining the speed of a polarization does \textit{not} dramatically affect the allowed values of a coupling constant. Hence, in this case, a tight constraint on the speed would not necessarily imply anything about the amplitude of the polarization (for the amplitudes of the different polarizations in this theory, see~\cite{Schumacher:2023cxh}). A similar argument applies to the scalar modes.

Einstein-\ae{}ther theory, therefore, provides examples where constraints on the speed both \textit{do} and \textit{do not} dramatically impact coupling constants 
This demonstrates that a small difference in propagation speed would not necessarily imply an undetectable amplitude.
        
\section{Conclusion}
\label{sec:conclusion}
In this work, we have proposed a novel method to search for additional polarizations that can propagate at different speeds. 
Observational evidence (the absence of gravitational Cherenkov radiation~\cite{Moore:2001bv, Kiyota:2015dla} and the joint detection of GWs and a gamma ray burst~\cite{LIGOScientific:2017zic}) implies that any scalar and vector polarizations must propagate at or faster than the speed of light, while the tensor polarizations propagate at the speed of light.
Therefore, any additional polarizations must arrive \textit{before} (or at best \textit{with}) the tensor modes. 
In light of this, we suggest starting from already detected events and searching back through earlier data for additional, non-tensorial polarizations with similar source parameters. 
This new method is necessary because traditional techniques of searching for simultaneous polarizations could entirely miss polarizations with even slightly different propagation speeds. 

The detection of additional polarizations in earlier data would be immediate evidence of new physics, but we must also consider the implications of no additional polarizations detected.
If they exist (as many modified theories of gravity suggest they might), additional polarizations may not be detected either because their amplitudes are not large enough for current detectors to measure or because they arrived before the portion of data searched. 
Therefore, a non-detection of additional polarizations in earlier data implies an upper bound on the amplitude for polarizations with detectable propagation speeds and a lower bound on the speed of propagation for polarizations with detectable amplitudes.  
We identify that closer events and longer observing times would allow us to impose more stringent constraints in both scenarios.
Finally, we point out that additional modes may appear to be isolated (i.e., not have tensor polarizations) if they arrive at dramatically different times, which can happen if their speed of propagation is even a tiny fraction larger than the speed of light. 

After introducing this new search technique, we focused on constraints that might be placed on the speed of propagation, exploring and addressing some of the limitations of our proposed method. 
We considered the challenges that arise because of limited observing time and gaps between different observing runs. 
Fortunately, collecting more data and ``stacking'' events resolves these issues.
We pointed out that the detectors' sensitivity changes across observing runs and propose ways to mitigate this effect (e.g., restricting the search to one run, lowering the SNR threshold, or only looking at sufficiently close events). 
Lastly, we examined existing constraints on the amplitude of additional polarizations and determined that these do not rule out the possibility of detection. 
We give an example of a specific theory to demonstrate how constraints on propagation speeds and the coupling constants of a theory (which also impact the amplitude) may be related. 

Even after considering all the above challenges, the method proposed here still seems to be a promising avenue for detecting or constraining additional, non-tensorial polarizations with different propagation speeds. 
This method could be immediately implemented with current GW templates in modified theories of gravity. 
For example, the Einstein-\ae{}ther waveform template from~\cite{Schumacher:2023cxh} could be easily adapted to this search (although, in this particular theory, estimates suggest that the improvement in constraints on the coupling constants would not be enough to justify this search, Sec.~\ref{sec:magnitude}).
The search is also dramatically simplified compared to a blind search because some parameters can be fixed (e.g., sky location and intrinsic parameters like chirp mass). 
However, we caution against attempting this search with GR templates. 
It is unclear (and highly unlikely) that the current GR template banks could pick up additional polarizations in the data, because the former can be dramatically functionally different from tensor polarizations (both geometrically and because of which harmonics dominate the signal~\cite{Chatziioannou:2012rf}). 
Even if GR templates can detect these additional polarizations, the source parameters may be highly biased, and the search will probably not be as efficient as if we had used more flexible templates. 
We leave the study of these questions to future work.

\acknowledgements
The authors would like to thank Katerina Chatziioannou, Carl-Johan Haster, Leah Jenks, Alexandria Tucker, and Tom Callister for helpful discussions. 
K.~S.~would like to acknowledge 
the National Science Foundation (NSF) Graduate Research Fellowship Program under Grant No. DGE – 1746047. 
N.~Y.~and K.~S.~acknowledge support from the Simons Foundation through Award No.~896696, the NSF through Grant No.~PHY-2207650 and NASA through Grant No.~80NSSC22K0806.
C.~T.~is supported by the Eric and Wendy Schmidt AI in Science Postdoctoral Fellowship, a Schmidt Sciences program.
D.E.H is supported by NSF grant  PHY-2110507, as well as by the Kavli Institute for Cosmological Physics through an endowment from the Kavli Foundation and its founder Fred Kavli.

This research has made use of data or software obtained from the Gravitational Wave Open Science Center (gwosc.org), a service of the LIGO Scientific Collaboration, the Virgo Collaboration, and KAGRA. This material is based upon work supported by NSF's LIGO Laboratory which is a major facility fully funded by the National Science Foundation, as well as the Science and Technology Facilities Council (STFC) of the United Kingdom, the Max-Planck-Society (MPS), and the State of Niedersachsen/Germany for support of the construction of Advanced LIGO and construction and operation of the GEO600 detector. Additional support for Advanced LIGO was provided by the Australian Research Council. Virgo is funded, through the European Gravitational Observatory (EGO), by the French Centre National de Recherche Scientifique (CNRS), the Italian Istituto Nazionale di Fisica Nucleare (INFN) and the Dutch Nikhef, with contributions by institutions from Belgium, Germany, Greece, Hungary, Ireland, Japan, Monaco, Poland, Portugal, Spain. KAGRA is supported by Ministry of Education, Culture, Sports, Science and Technology (MEXT), Japan Society for the Promotion of Science (JSPS) in Japan; National Research Foundation (NRF) and Ministry of Science and ICT (MSIT) in Korea; Academia Sinica (AS) and National Science and Technology Council (NSTC) in Taiwan.
This material is based upon work supported by NSF's LIGO Laboratory which is a major facility fully funded by the National Science Foundation.

\appendix
\section{Energy lost through additional polarizations}
\label{sec:appendix}
\renewcommand{\theequation}{\thesection.\arabic{equation}}
We here provide an estimate of the energy carried away by non-tensorial GW polarizations in modified theories. 
Starting from the general stress-energy tensor, Eq. 35.70 of~\cite{MTW}, 
\begin{align}
    T_{\mu\nu} &= \frac{1}{32\pi} \Big \langle (\nabla_\mu  {\h}_{\alpha\beta})(\nabla_\nu  {\h}^{\alpha\beta}) - \frac{1}{2} (\nabla_\mu  {\h})(\nabla_\nu  {\h}) \nonumber \\
    &\hspace{12mm} - 2 (\nabla_\beta  {\h}^{\alpha\beta})(\nabla_{(\nu}  {\h}_{\mu)\alpha}) \Big \rangle,
    \label{eqn:generalSET}
\end{align}
where $\h^{\mu\nu}$ is the trace reversed metric perturbation and the brackets $\langle \2\rangle$ represent a Brill-Hartle average~\cite{Isaacson:1968b, MTW}, averaging over several wavelengths. 
In GR, when $\h^{\mu\nu} = \h^{\mu\nu}_{TT}$, the second and third terms of Eq.~\eqref{eqn:generalSET} vanish, but $\h^{\mu\nu}$ may have additional (scalar and vector, non-TT) components in modified theories of gravity. 
Looking at the $T_{00}$ component of the above equation and expanding out the summations, we arrive at  
\begin{align}
    T_{00} &= \frac{1}{32\pi} \Big \langle  
    (\bar{\nabla}_0  {\h}_{jk})(\bar{\nabla}_0  {\h}^{jk}) - \frac{1}{2} (\bar{\nabla}_0  {\h})(\bar{\nabla}_0  {\h}) \nonumber \\
    &\hspace{12mm} -  (\bar{\nabla}_0  {\h}^{00})(\bar{\nabla}_{0}  {\h}_{00}) 
    - 2 (\bar{\nabla}_j  {\h}^{0j})(\bar{\nabla}_{0}  {\h}_{00}) \nonumber \\
    &\hspace{12mm}- 2 (\bar{\nabla}_k  {\h}^{jk})(\bar{\nabla}_{0}  {\h}_{0j})  
    \Big \rangle.
\end{align}
If we neglect terms of higher than second order in $\h$, then the covariant derivatives go to partial derivatives, and this becomes 
\begin{align}
    T_{00} &= \frac{1}{32\pi} \Big \langle 
    (\partial_0  {\h}_{jk})(\partial_0  {\h}^{jk})  - \frac{1}{2} (\partial_0  {\h})(\partial_0  {\h})\nonumber \\
    &\hspace{12mm} -  (\partial_0  {\h}^{00})(\partial_{0}  {\h}_{00}) 
    - 2 (\partial_j  {\h}^{0j})(\partial_{0}  {\h}_{00}) \nonumber \\
    &\hspace{12mm}- 2 (\partial_k  {\h}^{jk})(\partial_{0}  {\h}_{0j}) 
    \Big \rangle. \label{eqn:SET}
\end{align}
Following Eq.~(A.4) of~\cite{Schumacher:2023jxq}, the trace-reversed metric perturbation can be irreducibly decomposed as
\begin{subequations}
    \begin{align}
     \h^{00} &= C, \\
     \h^{0j} &= \frac{1}{v_S} N^j D + D^j_T, \\
     \h^{jk} &= \frac{\delta^{jk}}{3} A + \frac{1}{v_S^2} \left(N^j N^K - \frac{\delta^{jk}}{3} \right) B \nonumber \\
     &\hspace{4mm}+ \frac{2}{v_V} N^{(j} A^{k)}_T + A^{jk}_{TT},
    \end{align}
\end{subequations}
where $\partial_j D^j_T = 0 = \partial_j A^j_T$, $\partial_j A^{jk}_{TT} = 0 = \delta_{jk} A^{jk}_{TT}$, and $\mathbf{N}$ is a unit 3-vector aimed from the GW source to the detector.

Let us assume for simplicity that each deviation from GR scales like \textit{the same} small coupling constant, $\zeta$, and that we can neglect any terms of $\mathcal{O}(\zeta^2)$. 
In GR, the only non-vanishing piece of the GW is $A^{jk}_{TT}$, so every other function in this expression represents a deviation from GR that must thus scale with $\zeta$. 
Furthermore, the transverse-traceless polarizations may also be modified in beyond-GR theories of gravity. 
Explicitly writing $\zeta$ in as a bookkeeping parameter, the decomposition then becomes
\begin{subequations}
    \begin{align}
     \h^{00} &= \zeta C,\\
     \h^{0j} &= \frac{1}{v_S} N^j \zeta D + \zeta D^j_T, \\
     \h^{jk} &= \frac{\delta^{jk}}{3} \zeta A + \frac{1}{v_S^2} \left(N^j N^k - \frac{\delta^{jk}}{3} \right) \zeta B \nonumber \\
     &\hspace{4mm} + \frac{2}{v_V} N^{(j} \zeta A^{k)}_T + \zeta \delta A^{jk}_{TT} + A^{jk}_{TT}.
\end{align}
\end{subequations}
Now we can insert these expressions into Eq.~\eqref{eqn:SET} and analyze the result term by term. We will save the first term for last, as this is the interesting one. Starting with the second term, and pulling the constant $\zeta$ out of any derivatives, we have
\begin{equation}
    (\partial_0  {\h})(\partial_0  {\h}) = (\partial_0 [-\zeta C + \zeta A])^2 = \zeta^2 (\partial_0 [- C + A])^2= \mathcal{O}(\zeta^2)
\end{equation}
where we have used the fact that $\h = \eta_{\alpha\beta} \h^{\alpha\beta} = -\h^{00} + \h^{kk} = -\zeta C + \zeta A$
(since $A^{kk}_{TT} = 0$ and $N_k A^k_T = 0$).
Likewise,
\begin{equation}
    (\partial_0  {\h}^{00})(\partial_{0}  {\h}_{00}) = (\partial_0 [ \zeta C])^2 = \zeta^2 (\partial C)^2 = \mathcal{O}(\zeta)^2,
\end{equation}
and
\begin{align}
(\partial_j  {\h}^{0j})(\partial_{0}  {\h}_{00}) &= \zeta^2 \partial_j \left[\frac{1}{v_S} N^j D \right] \partial_0 \left[ C\right] = \mathcal{O}(\zeta^2),
\end{align}
(where we used $\partial_j D^j_T = 0$). 
Similarly, since $\partial_k A^{jk}_{TT} = 0$, we then have that
\begin{widetext}
\begin{align}
    (\partial_k  {\h}^{jk})(\partial_{0}  {\h}_{0j}) &= \partial_k  \left[\frac{\delta^{jk}}{3} \zeta A + \frac{1}{v_S^2} \left(N^j N^k - \frac{\delta^{jk}}{3} \right) \zeta B 
    + \frac{2}{v_V} N^{(j} \zeta A^{k)}_T + \zeta \delta A^{jk}_{TT} + A^{jk}_{TT} \right] 
    \partial_{0}  \left[-\frac{1}{v_S} N_j \zeta D - \zeta D_j^T \right], \nonumber \\
    &= \zeta^2 \partial_k  \left[\frac{\delta^{jk}}{3} A + \frac{1}{v_S^2} \left(N^j N^k - \frac{\delta^{jk}}{3} \right) B 
    + \frac{2}{v_V} N^{(j} A^{k)}_T \right] 
    \partial_{0}  \left[- \frac{1}{v_S} N_j  D - D_j^T \right] = \mathcal{O}(\zeta^2).
\end{align}
\end{widetext}
Thus, all of these terms can be neglected as they are of order $\mathcal{O}(\zeta^2)$.

Finally, let us turn to the first term of Eq.~\eqref{eqn:SET}. 
To simplify notation, we will use $x^0 = ct$ to write $\partial_0 = c^{-1} \partial_t$ and then use $\partial_t F = \dot{F}$ for any function $F$. Then, this first term becomes
\begin{align}
    \Big \langle (\partial_0  {\h}_{jk})(\partial_0  {\h}^{jk}) \Big \rangle&= \frac{1}{c^2} \Big \langle (\dot{A}^{jk}_{TT})^2 \nonumber \\
    & + 2
    \left[\frac{\delta_{jk}}{3} \zeta \dot{A} + \frac{1}{v_S^2} \left(N_j N_k - \frac{\delta_{jk}}{3} \right) \zeta \dot{B} 
    \right. \nonumber \\
    &\left.+ \frac{2}{v_V} N_{(j} \zeta \dot{A}^T_{k)} + \zeta \delta \dot{A}_{jk}^{TT} \right]  \dot{A}^{jk}_{TT} + \mathcal{O}(\zeta^2)
    \Big \rangle.
\end{align}
Recall that $N_j A^j_T = 0$ and $N_j A^{jk}_{TT} = 0 = \delta_{jk} A^{jk}_{TT}$. Therefore, the only part of this expression that does not vanish is 
\begin{align*}
    \Big \langle (\partial_0  {\h}_{jk})(\partial_0  {\h}^{jk}) \Big \rangle &= \frac{1}{c^2} \Big \langle 
    (\dot{A}_{jk}^{TT})^2  + 2 \zeta (\dot{A}^{jk}_{TT} \delta \dot{A}^{jk}_{TT} ) + \mathcal{O}(\zeta^2)
    \Big \rangle, 
\end{align*}
and we have 
\begin{equation}
    T_{00} = \frac{1}{32 c^2 \pi} \Big \langle (\dot{A}_{jk}^{TT})^2  + 2 \zeta (\dot{A}^{jk}_{TT} \delta \dot{A}^{jk}_{TT} ) + \mathcal{O}(\zeta^2)  \Big \rangle.
\end{equation}
Thus, we can see that energy carried away by additional, non-tensorial polarizations scales like $\zeta^2$ in this case and should not bias our measurement of the source parameters from analysis of the tensor polarizations (as long as $\zeta$ is small enough that we cannot detect a deviation from GR in the tensor polarizations, $\zeta^2$ is certainly negligible).

\section{Constraints on the propagation speed of scalar or vector polarizations: Leveraging multiple detectors}
\label{sec:independent}
The discussion in Sec.~\ref{sec:compare_to_tensors} hinges on searching for additional modes associated with the tensor modes from a particular event. 
However, if the difference in propagation speed is extreme, these different polarizations may arrive at dramatically different times and be hard to associate with each other. 
Therefore, it is worth considering how to constrain the speed of scalar or vector modes that seem to arrive independently of the tensor mode. 
To do this, we can take advantage of the difference in arrival times between different detectors.

\begin{figure}
    \centering
    \includegraphics[width=\linewidth]{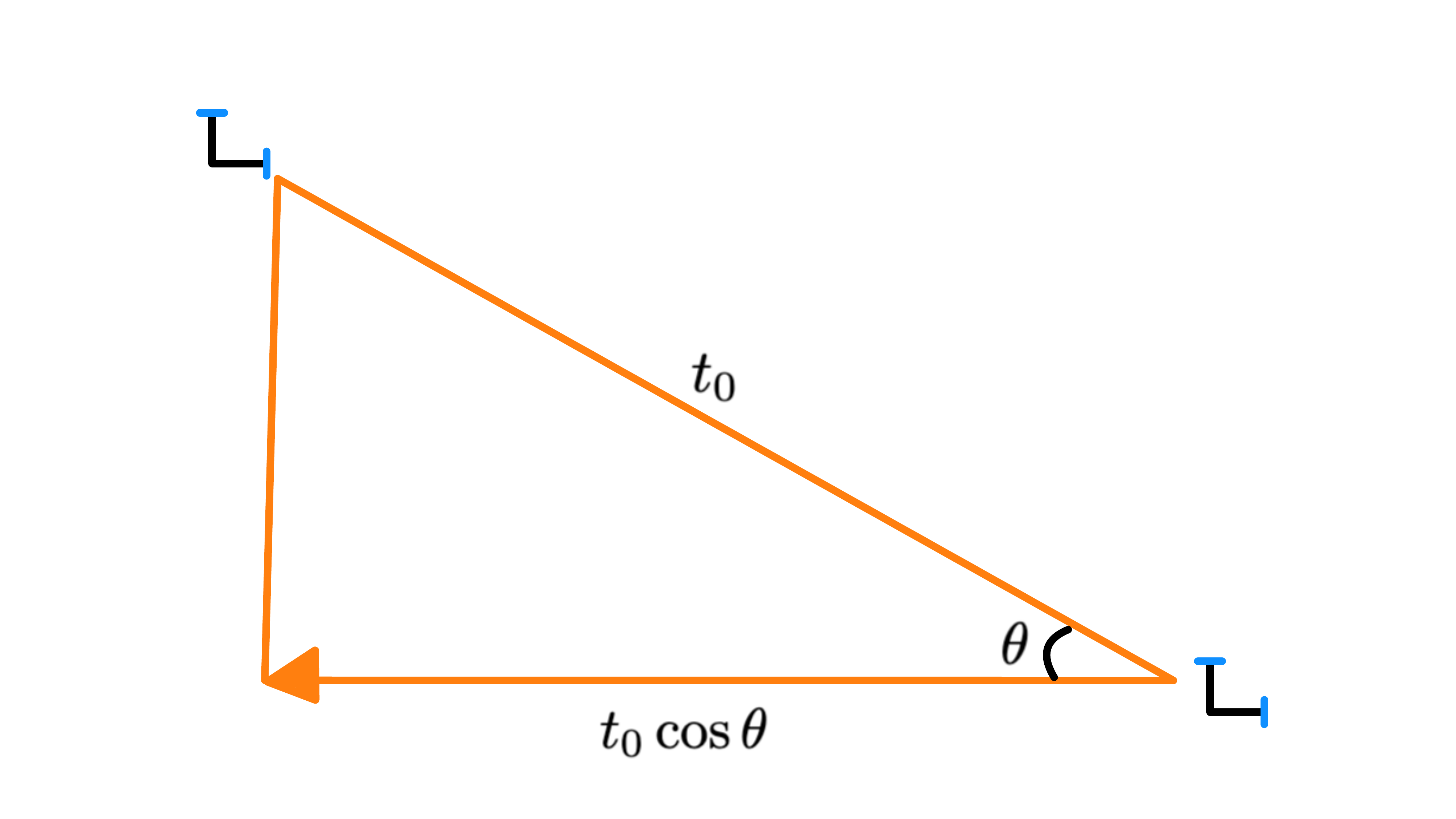}
    \caption{A sketch of two detectors separated by a light-travel time of $t_0 = d_{sep}/c$. An electromagnetic signal is propagating at an angle $\theta$ relative to the line connecting the detectors (line with arrow). From this sketch, it is easy to see that the time delay for a plane wave between the different detectors should be $t_0 \cos\theta$. }
    \label{fig:detectors_sketch}
\end{figure}

First let us review the methodology outlined by~\cite{Cornish:2017jml}, so that we can apply it to our particular problem. 
Consider two detectors which are separated by some distance $d_{\text{sep}}$. 
The amount of time it takes for light to travel on a straight line between the two sites is $t_0 = d_{\text{sep}}/c$. 
Now, if the light is propagating at some angle $\theta$ to the separation line between detectors, as sketched in Fig.~\ref{fig:detectors_sketch}, the time delay for electromagnetic signals between the different detectors will be $\tau_{\text{em}} = t_0\cos\theta$.
Assuming sources are roughly uniformly distributed throughout the sky, the time delays should be uniformly distributed between $-t_0 \leq \tau_{\text{em}} \leq t_0$. Thus, the likelihood of any particular time delay within this range of values is $p(\tau_{\text{em}}) = 1/(2t_0)$. 

The same logic can be applied to gravitational waves, as was done in~\cite{Cornish:2017jml} for the first three GW events that were detected. 
As described in Sec.~\ref{sec:compare_to_tensors}, these signals were most likely from tensor polarizations, which propagate with speed $v_T$ (at the time of writing of~\cite{Cornish:2017jml}, $v_T$ had not yet been constrained -- we will keep $v_T$  explicit in the equations below, in order to trace through the logic).
Thus, the travel time between the detectors is $d_{\text{sep}}/v_T = c \, t_0/v_T$ and the time delay for the tensor polarizations of GWs between two different detectors will be 
\begin{equation}
    \tau_T = \frac{ct_0}{v_T} \cos\theta = \frac{c}{v_T} \tau_{\text{em}}.
\end{equation}
Again, for isotropic sources, this would imply a uniform distribution of delay times between $-(c \, t_0/v_T)$ and $(c \, t_0/v_T)$, so the likelihood of any particular delay (given a speed $v_T$) is 
\begin{equation}
    p(\tau_T | v_T) = \begin{cases}
        \frac{v_T}{2 \, c \, t_0} &\text{for } -\frac{c \, t_0}{v_T} \leq \tau_T \leq  \frac{c \, t_0}{v_T}\\
        0 &\text{otherwise}
    \end{cases}
\end{equation}
However, even if the sources are uniformly distributed, the angular pattern functions (described in ~\cite{PWGravity2014}) imply that the response of the detector is not the same at all angles. 
Thus, this distribution of delay times must be weighted by the network power pattern, summed over $k$ detectors, as given in~\cite{Schutz:2011tw}:
\begin{equation}
    P_T(\theta, \phi) = \sum_{k} F_{+,k}^2(\theta, \phi) + F_{\times, k}^2(\theta, \phi)\,,
\end{equation}
where ($\theta,\phi$) are sky angles.
Reference~\cite{Cornish:2017jml} then used Bayes' theorem to arrive at the posterior distribution for $v_T$: 
\begin{equation}
    p(v_T | \tau_T) = \frac{p(\tau_T | v_T) p(v_T)}{p(\tau_T)}
\end{equation}
and used numerical sampling methods to determine the shape of this posterior from the three events they considered and to place constraints on $v_T$.

Note that the constraints already placed by~\cite{Cornish:2017jml} do not apply to additional, non-tensorial modes because they were placed with tensor polarizations. 
To place a similar constraint for non-tensorial modes, \textit{the analysis would have to be repeated after a conclusive detection of additional polarizations}. 
We would be able to send $v_T \rightarrow v_S$ or $v_T \rightarrow v_V$ depending on the polarization detected, and update the network power pattern to be either 
\begin{align*}
    P_S(\theta, \phi) &= \sum_{k}  F_{b,k}^2(\theta, \phi) + F_{L, k}^2(\theta, \phi)\,, \\
    &\text{or} \\
    P_V(\theta, \phi) &= \sum_{k}  F_{X,k}^2(\theta, \phi) + F_{Y, k}^2(\theta, \phi)\,.
\end{align*}
respectively.
If additional polarizations \textit{were} detected without corresponding tensor modes (i.e., in isolation), this method might provide an upper bound on their speed, because $\tau_{S,V} \leq c \, t_0/v_{S,V} \Rightarrow v_{S,V} \leq c \,  t_0/\tau_{S,V}$, so the larger the value of $\tau_{S,V}$, the smaller $v_{S,V}$ has to be. Intuitively, slower waves will take longer to travel between detectors, so large time delays set upper bounds on the speed.
An upper bound would allow us to calculate the maximum difference in arrival times between non-tensorial polarizations and tensorial polarizations from the same event.
This could potentially be used to narrow the region of data, or time-window, that should be searched for corresponding polarizations, although it requires combining the constraints from many different detections, since we would not know from any individual event if the delay in arrival times was due to speeds or angles.

This method hinges on observing the same polarization ``in coincidence'' in multiple detectors. 
For example, standard LIGO searches exclude signals with Hanford-Livingston time delays greater than 15 ms~\cite{LIGOScientific:2016vbw}.
If the propagation speeds were much less than the speed of light, it might take too long for these polarizations to travel from one detector to the other for them to still be recognized as the same event. 
Fortunately, since $v_{S, V} \geq c$, the travel time of additional polarizations should always be less than that of light, and it should be simple to correlate events between detectors for any additional polarizations.

On the other end, it is possible that additional polarizations might travel so fast that LIGO would see their arrival in multiple detectors as simultaneous events and would not be able to reveal more about their speeds. LIGO's timing resolution is 1$\mu$s~\cite{Sullivan:2023cqg}. Thus, to be able to distinguish between different speeds, let us conservatively estimate that we would need a difference in arrival times between detectors of at least 2 $\mu$s. Then, $\tau_{S,V} \geq 2 \2 \mu$s $\Rightarrow v_{S,V} \leq c \, t_0/2$ $\mu$s. For a separation between detectors of 3000 km (as between LIGO Hanford and LIGO Livingston), this means $t_0 \approx 10$ ms and hence $v_{S,V} \leq 5000c$ would be distinguishable. This is a ridiculously large speed, so we consider this method still potentially useful.

Since additional polarizations have not yet been observed, we put this method aside for the moment to be used if such polarizations are detected.

\bibliography{main, notInspire}

\end{document}